\documentclass[times,twocolumn]{aastex62}

\usepackage{CJK}
\usepackage{amsmath}
\usepackage{multirow}
\usepackage{cases}
\usepackage{graphicx}
\usepackage{subfigure}
\usepackage{natbib}
\usepackage{color}
\usepackage{tabularx}
\usepackage{bm}
\usepackage{threeparttable}
\usepackage{gensymb}

\defcitealias{OB110950}{C12}
\defcitealias{OB110950_AO}{T22}
\defcitealias{Three_planet_candidates}{Z25}

\newcommand{\thetae}{\theta_{\rm E}}

\newcommand{\murelGeo}{\mu_{\rm rel, G}}

\newcommand{\te}{t_{\rm E}}

\newcommand{\event}{OGLE-2011-BLG-0950}

\DeclareGraphicsExtensions{.pdf,.png,.jpg}

\shorttitle{}
\shortauthors{Zhang et al.}

\begin{document}
\begin{CJK*}{UTF8}{gbsn}
\title{{\large A Newly Identified Degeneracy Keeps the Planetary Interpretation Viable for OGLE-2011-BLG-0950}}

\correspondingauthor{Jiyuan Zhang, Weicheng Zang}
\email{zhangjy22@mails.tsinghua.edu.cn, zangweicheng@westlake.edu.cn}

\author[0000-0002-1279-0666]{Jiyuan Zhang}
\affiliation{Department of Astronomy, Tsinghua University, Beijing 100084, China}
%zhangjy22@mails.tsinghua.edu.cn

\author[0000-0001-6000-3463]{Weicheng Zang}
\affiliation{Department of Astronomy, Westlake University, Hangzhou 310030, Zhejiang Province, China}

\author[0000-0001-5207-5619]{Andrzej Udalski}
\affiliation{Astronomical Observatory, University of Warsaw, Al. Ujazdowskie 4, 00-478 Warszawa, Poland}
%udalski@astrouw.edu.pl

\author[0000-0003-0626-8465]{Hongjing Yang}
\affiliation{Westlake Institute for Advanced Study, Hangzhou 310030, Zhejiang Province, China}
\affiliation{Department of Astronomy, Westlake University, Hangzhou 310030, Zhejiang Province, China}
%yanghongjing@westlake.edu.cn

\author[0000-0001-8317-2788]{Shude Mao}
\affiliation{Department of Astronomy, Westlake University, Hangzhou 310030, Zhejiang Province, China}

\author[0000-0002-0548-8995]{Micha{\l}~K. Szyma\'{n}ski}
\affiliation{Astronomical Observatory, University of Warsaw, Al. Ujazdowskie 4, 00-478 Warszawa, Poland}
%msz@astrouw.edu.pl

\author[0000-0002-7777-0842]{Igor Soszy\'{n}ski}
\affiliation{Astronomical Observatory, University of Warsaw, Al. Ujazdowskie 4, 00-478 Warszawa, Poland}
%soszynsk@astrouw.edu.pl

\author[0000-0002-9245-6368]{Radoslaw Poleski}
\affiliation{Astronomical Observatory, University of Warsaw, Al. Ujazdowskie 4, 00-478 Warszawa, Poland}
%rpoleski@astrouw.edu.pl

\author[0000-0001-6364-408X]{Krzysztof Ulaczyk}
\affiliation{Department of Physics, University of Warwick, Gibbet Hill Road, Coventry, CV4~7AL,~UK}
%kulaczyk@astrouw.edu.pl

\author[0000-0002-2339-5899]{Pawe{\l} Pietrukowicz}
\affiliation{Astronomical Observatory, University of Warsaw, Al. Ujazdowskie 4, 00-478 Warszawa, Poland}
%pietruk@astrouw.edu.pl

\author[0000-0003-4084-880X]{Szymon Koz{\l}owski}
\affiliation{Astronomical Observatory, University of Warsaw, Al. Ujazdowskie 4, 00-478 Warszawa, Poland}
%simkoz@astrouw.edu.pl

\author[0000-0002-2335-1730]{Jan Skowron}
\affiliation{Astronomical Observatory, University of Warsaw, Al. Ujazdowskie 4, 00-478 Warszawa, Poland}
%jskowron@astrouw.edu.pl

\author[0000-0001-7016-1692]{Przemek Mr\'{o}z}
\affiliation{Astronomical Observatory, University of Warsaw, Al. Ujazdowskie 4, 00-478 Warszawa, Poland}
%pmroz@caltech.edu

\author[0000-0002-5029-3257]{Sean Terry}
\affiliation{Code 667, NASA Goddard Space Flight Center, Greenbelt, MD 20771, USA}
\affiliation{Department of Astronomy, University of Maryland, College Park, MD 20742, USA}

\author{Andrew Gould}
\affiliation{Max-Planck-Institute for Astronomy, K\"onigstuhl 17, 69117 Heidelberg, Germany}
\affiliation{Department of Astronomy, Ohio State University, 140 W. 18th Ave., Columbus, OH 43210, USA}
%\email{gould.34@osu.edu}

%shude.mao@gmail.com

\begin{abstract}
The microlensing event OGLE-2011-BLG-0950 exhibits the well-known ``Planet/Binary'' degeneracy, in which distinct lens configurations produce similar light curves but imply substantially different mass ratios between the lens components. A previous study suggested that high-resolution imaging could break this degeneracy through differences in the lens-source relative proper motion. In this work, we identify a new planetary model for this event that arises from a newly identified degeneracy, simultaneously reproducing the observed light curve and remaining consistent with the relative proper motion measured from high-resolution imaging. By combining constraints from the light-curve modeling and high-resolution observations, we infer a lens system consisting of a $\sim 1~M_{\odot}$ host star orbited by a $\sim 1.5~M_{\rm Jup}$ planet, with a projected separation of about 2 or 8 au, subject to the ``Close/Wide'' degeneracy. Our reanalysis of the color-magnitude diagram further indicates that the source star has unresolved companions that contribute non-negligible blended light, highlighting the importance of carefully accounting for source and lens companions in future Roman microlensing analyses. Finally, we show that adopting a single mass--luminosity relation significantly underestimates the uncertainties in the inferred lens properties for host masses $\gtrsim 1~M_{\odot}$.
\end{abstract}

\section{Introduction}\label{sec:intro}

The gravitational microlensing technique is most sensitive to planets located near or beyond Jupiter-like orbital separations \citep{Shude1991,Andy1992,BennettRhie}. To date, six homogeneous planet samples have been produced by microlensing surveys \citep{mufun,Cassan2012,Suzuki2016,Wise,OGLE_wide,OB160007}. Among these, \cite{mufun} provided the first measurement of the planet occurrence rate from microlensing observations. In addition, \cite{OB160007} presented the largest sample to date, comprising 63 planets, and identified a bimodal distribution in the planet-to-host mass ratio ($q$), with two distinct populations corresponding to super-Earths/Neptune-like planets and gas giants.

Degeneracies are common in microlensing planet detections. Some degeneracies, such as the ``Close/Wide'' degeneracy, for which two planetary solutions are approximately related by $s \leftrightarrow s^{-1}$ \citep{Griest1998,Dominik1999,An2005}, where $s$ is the projected separation between the planet and host in units of the Einstein radius, and the ``inner/outer'' degeneracy \citep{GG1997}, for which the source trajectory passes either inside (``inner'') or outside (``outer'') the planetary caustic(s) relative to the central caustic, generally yield similar values of the mass ratio and therefore do not significantly affect statistical studies of the planetary mass-ratio function. However, there are two types of degeneracy that frequently arise in planetary events that can introduce substantial ambiguities. 

The first is the binary-lens single-source (2L1S) versus single-lens binary-source (1L2S) degeneracy \citep{Gaudi1998}. This degeneracy typically arises in events exhibiting bump-like anomalies, which can also be reproduced by a 1L2S configuration for which a much fainter secondary source passes closer to the lens than the primary source. Three approaches are commonly employed to discriminate between the two interpretations. First, one can compare the goodness of fit by examining the $\chi^{2}$ difference between the 2L1S and 1L2S models. Second, the two-source scenario generally predicts a color difference between the primary and secondary sources due to their differing luminosities \citep{Shude1991,Gaudi1998}. Third, if finite-source effects are detected for the secondary source and imply an unrealistically low lens-source relative proper motion, the 1L2S model can be rejected. Nevertheless, in many cases the 1L2S interpretation cannot be definitively excluded. For example, during the construction of the statistical sample by \cite{OB160007}, nine candidate planetary events were excluded because of the unresolved 2L1S/1L2S degeneracy.

The second is the ``Planet/Binary'' degeneracy \citep{HanGaudi2008}, which frequently arises in high-magnification events and can lead to substantially different inferred values of $q$, spanning both the planetary and stellar-binary regimes. In the analysis of \cite{OB160007}, three events affected by this degeneracy were therefore excluded from the mass-ratio function study. 
In some cases, the planetary interpretation yields a well-constrained measurement of the lens-source relative proper motion, $\mu_{\rm rel}$, whereas the stellar-binary model provides only a weak lower limit on $\mu_{\rm rel}$ (see the summary in \citealt{Three_planet_candidates}, hereafter \citetalias{Three_planet_candidates}). High-resolution imaging obtained with space-based telescopes or ground-based adaptive optics (AO) facilities can sometimes discriminate between the two scenarios. If the lens and source are resolved and the measured separation implies a $\mu_{\rm rel}$ that is inconsistent with the planetary models, the planetary interpretation can be rejected. This approach was demonstrated by \cite{OB110950_AO} (hereafter \citetalias{OB110950_AO}) for OGLE-2011-BLG-0950, for which Keck AO imaging yielded a heliocentric lens-source relative proper motion that, after transformation to the geocentric frame, corresponds to $\mu_{\rm rel,G} = 4.06 \pm 0.22~{\rm mas~yr^{-1}}$, in clear disagreement with $\mu_{\rm rel,G} = 1.05 \pm 0.20~{\rm mas~yr^{-1}}$ predicted by the planetary model based on their light curve and color-magnitude diagram analysis. 

Recently, \citetalias{Three_planet_candidates} identified a new ``Planet Finite/Planet Point'' degeneracy in events showing the ``Planet/Binary'' degeneracy. 
The ``Planet Finite'' solutions have well-detected finite-source effects and therefore yield tightly constrained $\mu_{\rm rel}$ that can be strongly tested with high-resolution imaging, whereas the degenerate ``Planet Point'' solutions, similar to the stellar-binary models, exhibit weak constraints on finite-source effects and, consequently, on the relative proper motion. 
As a result, the ``Planet Point'' branch generally cannot be ruled out by high-resolution imaging. 
Motivated by this newly identified degeneracy, we re-modeled the light curve of OGLE-2011-BLG-0950 and confirmed that such ``Planet Point'' solutions are also present in this event, in addition to the previously published planetary solution that corresponds to the ``Planet Finite'' branch and is ruled out by the Keck AO relative proper motion constraint, thereby keeping the planetary interpretation viable. 
In Section~\ref{sec:model}, we describe the light-curve analysis of this event. In Sections~\ref{sec:source} and \ref{sec:lens}, we present a detailed assessment of the source properties and derive the lens properties. Finally, in Section~\ref{sec:dis}, we discuss the broader implications and potential directions for future work.

\section{Light-curve Analysis}\label{sec:model}

\begin{figure}
    \centering
    \includegraphics[width=\columnwidth]{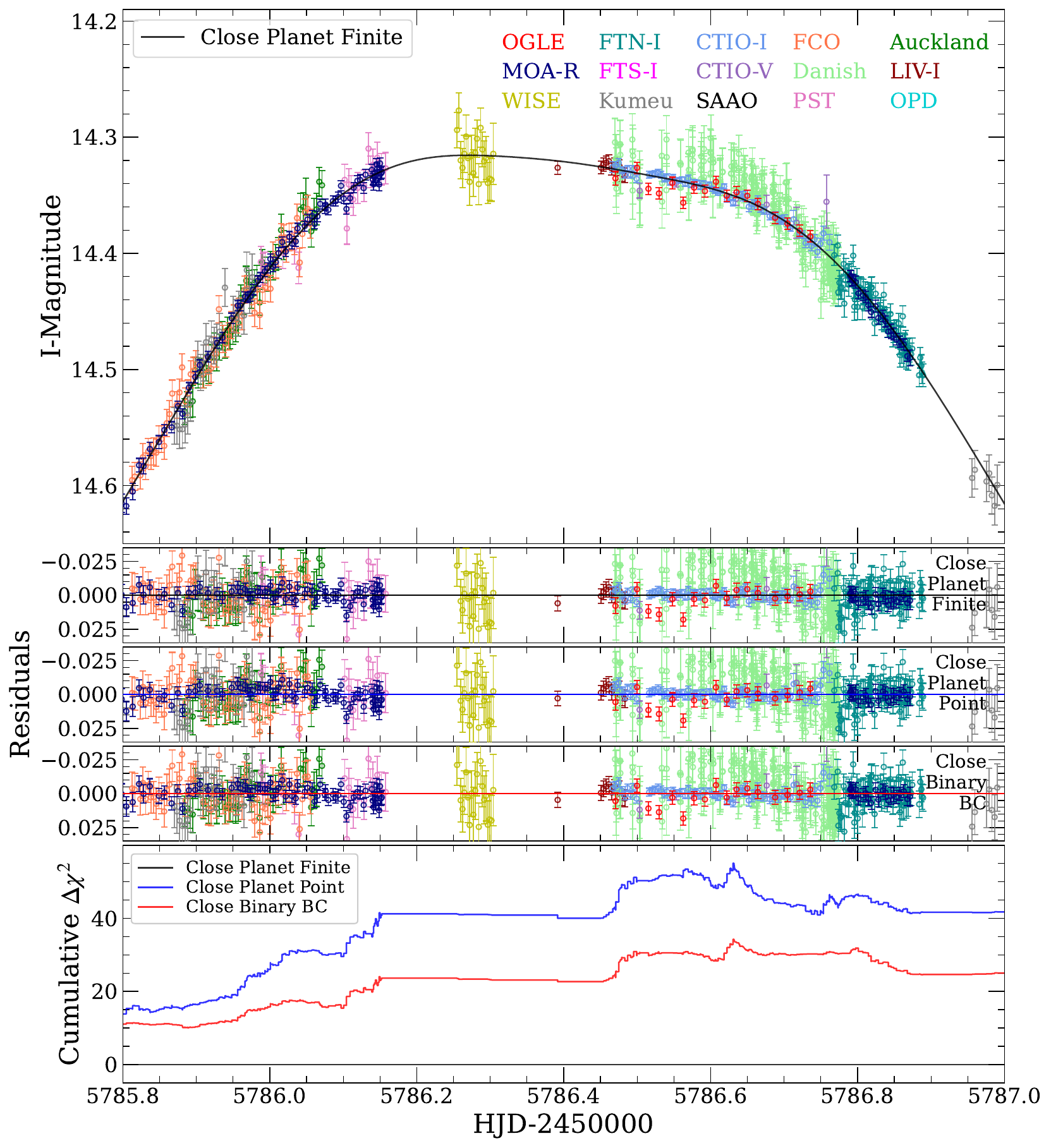}
    \caption{Observed light curves and 2L1S lensing models for \event. The various data sets are shown in different colors. The asymmetric peak is explained by 2L1S models. The black curve in the top panel represents the ``Close Planet Finite'' solution, which yields the lowest $\chi^2$ among the tested 2L1S models. Residuals for the best-fit ``Planet Finite'', ``Planet Point'', and ``Binary'' solutions are shown in the second through fourth panels, respectively. The bottom panel presents the cumulative $\Delta\chi^2$ distributions of these solutions relative to the ``Close Planet Finite'' solution. The lensing parameters for all 2L1S solutions are listed in Table~\ref{OB110950_parm_2L1Sstatic}.
}
    \label{fig:lc}
\end{figure}

Figure~\ref{fig:lc} displays the observed light curve and the best-fit 2L1S models for \event \footnote{The astute reader will note that, compared to \citetalias{OB110950_AO}, the data in the top panel of Figure~\ref{fig:lc} are flipped with respect to the model: epochs where the data lie brighter than the model in our plot lie fainter than the model in \citetalias{OB110950_AO} (and vice versa). 
We have verified that our plot preserves the intended data--model relation implied by the publicly available data; interested readers can independently verify this using those data. 
Similar plotting inversions have appeared in the literature, e.g., Figure~2 of \cite{OB161195_MOA}, as noted by \cite{OB161195_Gould}.}. The data used in our analysis are the same as those adopted by \citetalias{OB110950_AO}, except for the OGLE data, for which we re-reduced the OGLE-IV $I$-band data set. The OGLE-IV $V$-band data are used in Section~\ref{sec:source} to measure the source color. The reported $I$- and $V$-band magnitudes have been calibrated onto the OGLE-III photometric system \citep{OGLEIII} using common nearby field stars observed in both OGLE-IV and OGLE-III. The photometric uncertainties were rescaled following \citet{MB11293} to ensure that the reduced $\chi^2$ for each data set is unity.

A standard 2L1S model is characterized by seven parameters. In addition to the mass ratio $q$ and the projected separation $s$, it includes the Paczy\'{n}ski parameters \citep{Paczynski1986}: $t_0$, $u_0$, and $\te$, which correspond to the time of closest approach of the source to a reference point in the lens system, the impact parameter in units of the angular Einstein radius $\thetae$, and the Einstein-radius crossing time, respectively. We adopt the magnification center as the reference point, using the definition in \citetalias{Three_planet_candidates}. The sixth parameter, $\rho$, represents the angular source radius $\theta_*$ normalized by $\thetae$ ($\rho = \theta_*/\thetae$). The final parameter, $\alpha$, is the angle between the source trajectory and the binary axis. For each data set $i$, we introduce two flux parameters, $f_{{\rm S},i}$ and $f_{{\rm B},i}$, to represent the source flux and blended flux, respectively. 

We adopt the approach of \citetalias{Three_planet_candidates} to perform the 2L1S modeling, starting with a grid search and subsequently refining the parameters using the \texttt{emcee} ensemble sampler \citep{emcee} and the Nelder-Mead simplex algorithm \footnote{Implemented via \texttt{scipy.optimize.fmin}. See \url{https://docs.scipy.org/doc/scipy/reference/generated/scipy.optimize.fmin.html\#scipy.optimize.fmin}}. We compute the 2L1S magnification using the most recent release of the advanced contour integration code \texttt{VBMicrolensing} \citep{Bozza2010, Bozza2018, VBMicrolensing2025}. 

Neither the original discovery paper (\citealt{OB110950}; hereafter \citetalias{OB110950}) nor \citetalias{OB110950_AO} explicitly stated whether limb-darkening effects for the source star were considered in the modeling. In our analysis, we explicitly include limb darkening. Based on the source properties in Section~\ref{sec:source}, we adopt linear limb-darkening coefficients from \citet{Claret2011} appropriate for each observational bandpass.

\begin{figure}
    \centering
    % \captionsetup{type=figure}
    \includegraphics[width=\columnwidth]{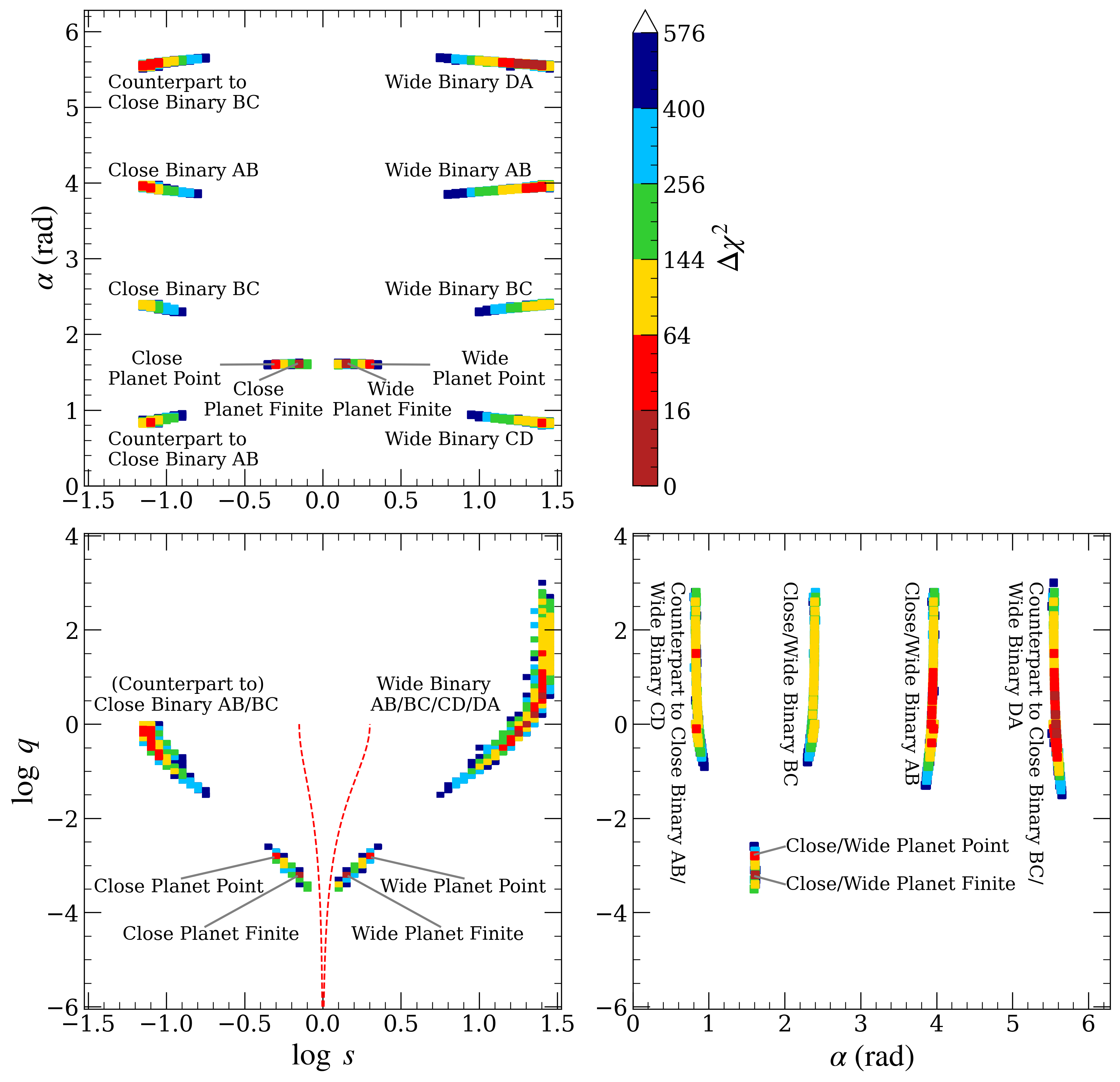}
    \caption{The $\chi^2$ landscape in the $(\log s, \log q, \alpha)$ parameter space as explored by a grid search, following the approach described by \citetalias{Three_planet_candidates}. Color coding is used to represent the statistical significance of grid points: dark red, red, yellow, green, blue, and dark blue correspond to deviations within $1n\sigma$, $2n\sigma$, $3n\sigma$, $4n\sigma$, $5n\sigma$, and $6n\sigma$, respectively, where $n = 4$. Grid points exceeding the $6n\sigma$ threshold are not shown. A total of 12 local minima are identified and labeled, two of which are redundant ``counterpart'' minima of other ``Binary'' solutions (i.e., intrinsically identical under the $q \leftrightarrow 1/q$ symmetry). 
    % two of which correspond to ``Binary'' solutions' counterpart. 
    The two red dashed lines mark the boundaries separating resonant and non-resonant caustic regimes, as defined by Equations (60) and (61) of \citet{Dominik1999}.}
    \label{OB110950_gridsearch}
\end{figure}

\begin{figure}
    %\centering
    \includegraphics[width=0.47\textwidth]{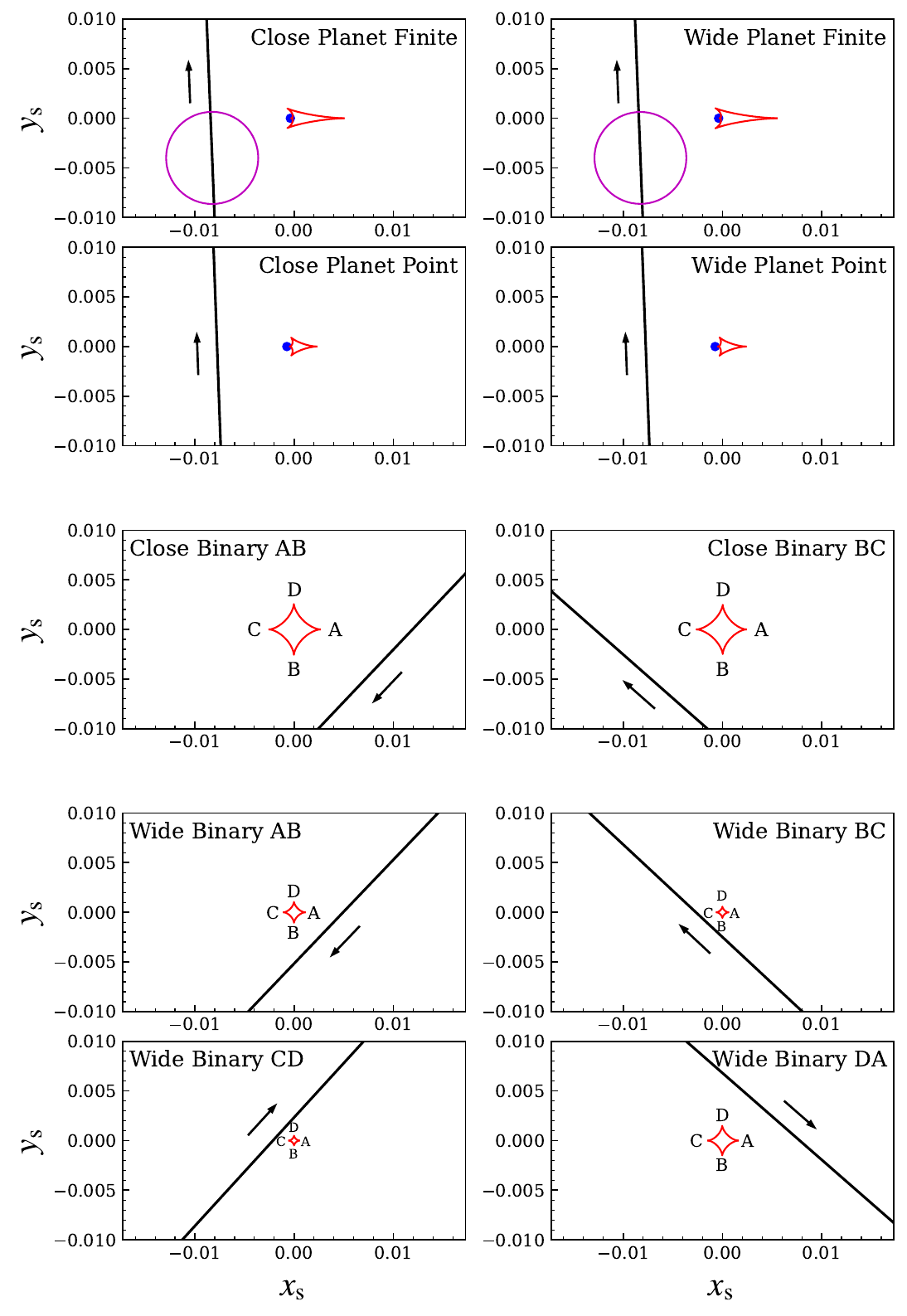}
    \caption{Caustic configurations for the ten 2L1S solutions listed in Table~\ref{OB110950_parm_2L1Sstatic}. In each panel, the caustic structure is shown by the red curves, the position of the primary lens is indicated by a blue dot, and the source-lens relative trajectory is plotted as a black line, with an arrow denoting the direction of source motion. For the two ``Planet Finite'' solutions, finite-source effects are detected, and the magenta circles indicate the angular radius of the source. For the six ``Binary'' solutions, individual caustic cusps are labeled as ``A'', ``B'', ``C'', and ``D'', according to the naming convention introduced by \citetalias{Three_planet_candidates}.}
\label{OB110950_caustic}
\end{figure}

Figure \ref{OB110950_gridsearch} presents the results of the grid search. Similar to the three events affected by the ``Planet/Binary'' degeneracy analyzed by \citetalias{Three_planet_candidates}, \event\ exhibits six ``Binary'' solutions. 
% Following the definitions of \citetalias{Three_planet_candidates}, we label these solutions as 
Following the nomenclature of \citetalias{Three_planet_candidates}, we label the four cusps of the relevant diamond-shaped caustic as A--D and denote each solution by the consecutive cusp pair traversed by the source trajectory; we therefore label these solutions as ``Close Binary AB'', ``Close Binary BC'', ``Wide Binary AB'', ``Wide Binary BC'', ``Wide Binary CD'', and ``Wide Binary DA''. The corresponding caustic structures and source trajectories are shown in Figure \ref{OB110950_caustic}. We note that both \citetalias{OB110950} and \citetalias{OB110950_AO} reported only two ``Binary'' solutions, which are the ``Close Binary BC'' and ``Wide Binary DA'' in our solutions.

\begin{figure}
    \centering
    % \captionsetup{type=figure}
    \includegraphics[width=\columnwidth]{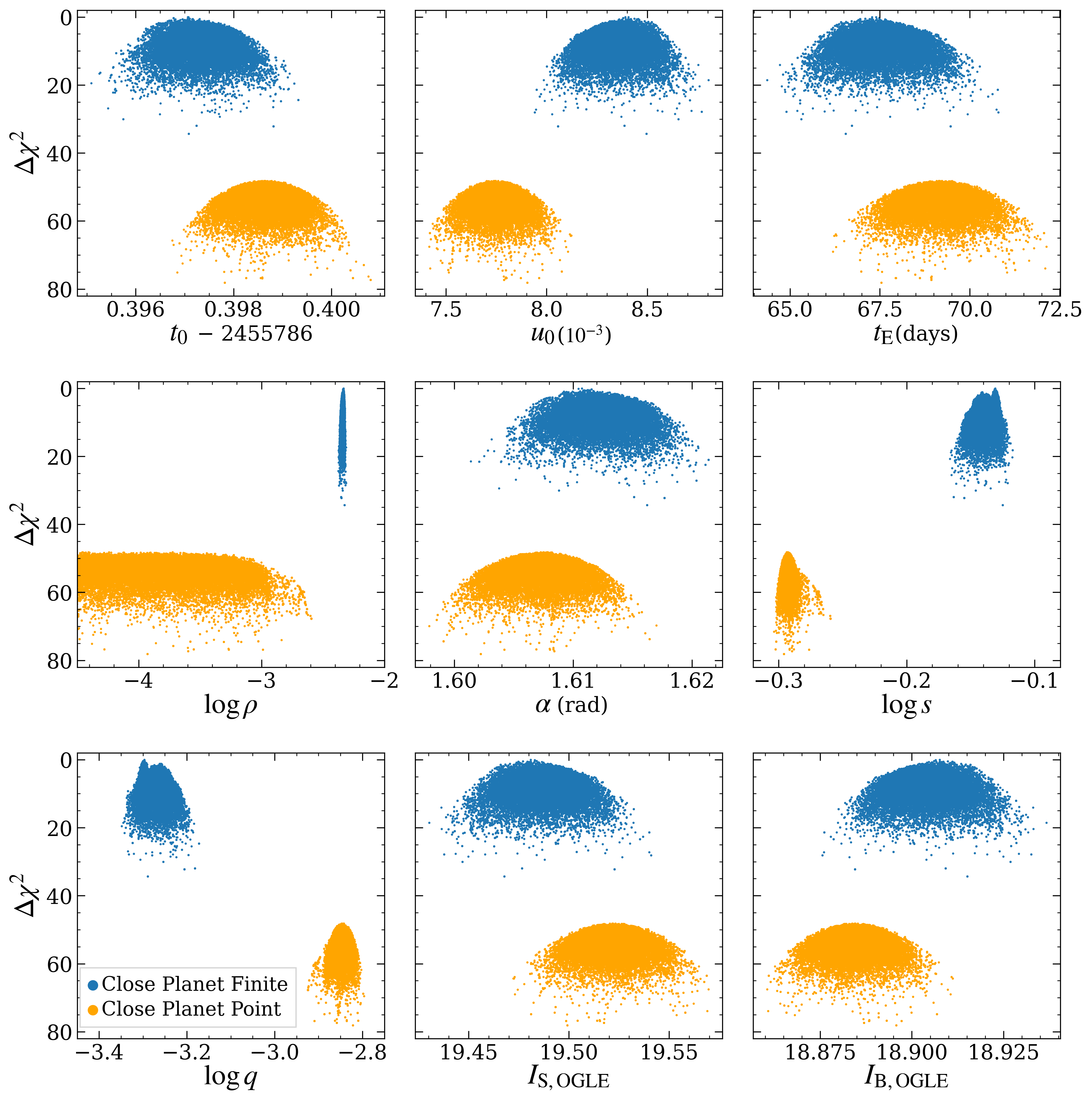}
    \caption{Combined MCMC chains for the ``Close Planet Finite'' (blue points) and ``Close Planet Point'' (orange points) solutions. The nine panels display the seven 2L1S parameters, as well as the source flux and blend flux in the OGLE-III magnitude system. The y-axis shows the $\Delta\chi^2$ of each point relative to the minimum $\chi^2$ of the combined chains.}
    \label{OB110950_combined_MCMC_chain}
\end{figure}

For the planetary solutions, analogous to KMT-2022-BLG-0954 in \citetalias{Three_planet_candidates}, we identify two pairs of solutions: one pair without finite-source effects (``Close Planet Point'' and ``Wide Planet Point'') and one pair with finite-source effects (``Close Planet Finite'' and ``Wide Planet Finite''). Among these, the ``Planet Finite'' solutions were the ones reported by \citetalias{OB110950} and \citetalias{OB110950_AO}. The ``Planet Point'' solutions were newly discovered. Figure \ref{OB110950_combined_MCMC_chain} shows the combined MCMC chains for ``Close Planet Finite'' and ``Close Planet Point'', demonstrating that the two solutions are indeed distinct. 

Table \ref{OB110950_parm_2L1Sstatic} lists the 2L1S parameters for all solutions, as obtained from the MCMC. The ``Planet Finite'' solutions provide the best fit to the light curves. Among the ``Binary'' solutions, ``Close Binary BC'' and ``Wide Binary DA'' have the lowest $\chi^2$, but they are still disfavored relative to the ``Planet Finite'' solutions by $\Delta\chi^2 = 26$, which is consistent with the $\chi^2$ difference reported by \citetalias{OB110950_AO}. The remaining ``Binary'' solutions are even more disfavored, with $\Delta\chi^2$ ranging from 39 to 61. 

The ``Planet Point'' solutions are disfavored by $\Delta\chi^2 = 48$, which appears statistically significant, but does not necessarily imply that they can be ruled out. Although the ``Planet Finite'' solutions are favored by $\Delta\chi^2 \geq 26$ relative to any other solutions, they are already excluded by the Keck AO observations reported by \citetalias{OB110950_AO}, which almost certainly indicates the presence of systematics in the light curves. As shown in the cumulative $\Delta\chi^2$ plot in Figure \ref{fig:lc}, the $\Delta\chi^2$ contributions for the ``Close Binary BC'' and ``Close Planet Point'' solutions originate from nearly the same regions, suggesting that both are affected by similar systematics. Therefore, the ``Planet Point'' solutions remain viable for this event.

\begin{table*}
%[htb]
    \renewcommand\arraystretch{1.15}
    \centering
    \caption{2L1S Parameters for OGLE-2011-BLG-0950}
    %\caption{Lensing Parameters for KMT-2022-BLG-0440}
    
    % \resizebox{\textwidth}{!}{
    
    \begin{tabular}{c|c|c|r r r r r r r r}
    \hline
    \hline
    \multicolumn{2}{c|}{Model} & 
    $\chi^2$/dof & 
    $t_{0}$ (${\rm HJD}^{\prime}$) &
    $u_{0}~(10^{-3})$ &
    $\te$ (days) &
    $\rho~(10^{-3})$ &
    $\alpha$ (rad) &
    $s$ &
    $\log q$ &
    $I_{\rm S, OGLE}$ \\
    
    \hline
    \multirow{8}{*}{Close} 
    
    & \textbf{Planet Finite}
    & $\boldsymbol{6992.3/6993}$

    & $\boldsymbol{5786.3973}$ & $\boldsymbol{8.35}$ & $\boldsymbol{67.56}$ & $\boldsymbol{4.56}$ & $\boldsymbol{1.612}$ & $\boldsymbol{0.727}$ & $\boldsymbol{-3.271}$ & $\boldsymbol{19.486}$ \\
    
    & & & $\boldsymbol{0.0005}$ & $\boldsymbol{0.11}$ & $\boldsymbol{0.80}$ & $\boldsymbol{0.08}$ & $\boldsymbol{0.002}$ & $\boldsymbol{0.011}$ & $\boldsymbol{0.024}$ & $\boldsymbol{0.014}$ \\ %updated 
    %the bold is larger in font than unbold, should be fixed

    \cline{2-11}
    & {Planet Point}
    & $7040.5/6993$
    & $5786.3987$ & $7.75$ & $69.18$ & $<1.91$ & $1.607$ & $0.511$ & $-2.849$ & $19.522$ \\
    & & & $0.0005$ & $0.09$ & $0.74$ & $-$ & $0.002$ & $0.006$ & $0.016$ & $0.012$ \\ %updated 

    \cline{2-11}
    & Binary AB 
    & $7031.3/6993$
    & $5786.3972$ & $8.65$ & $64.46$ & $<2.69$ & $3.954$ & $0.073$ & $-0.181$ & $19.423$ \\
    & & & $0.0008$ & $0.10$ & $0.69$ & $-$ & $0.005$ & $0.001$ & $0.059$ & $0.013$ \\ %updated 

    \cline{2-11}
    & {Binary BC} 
    & $7018.3/6993$
    & $5786.3974$ & $8.51$ & $65.20$ & $<2.63$ & $2.419$ & $0.075$ & $0.314$ & $19.436$ \\
    & & & $0.0006$ & $0.11$ & $0.75$ & $-$ & $0.006$ & $0.002$ & $0.068$ & $0.013$ \\ %updated 
    
    \hline
    \multirow{12}{*}{Wide} 
    & {Planet Finite}
    & $6992.4/6993$

    & $5786.3973$ & $8.35$ & $67.60$ & $4.58$ & $1.612$ & $1.365$ & $-3.270$ & $19.487$ \\
    
    & & & $0.0005$ & $0.11$ & $0.81$ & $0.08$ & $0.002$ & $0.021$ & $0.024$ & $0.014$ \\ %updated 

    \cline{2-11}
    & {Planet Point}
    & $7040.6/6993$
    & $5786.3987$ & $7.74$ & $69.27$ & $<1.95$ & $1.607$ & $1.949$ & $-2.845$ & $19.523$ \\
    & & & $0.0005$ & $0.10$ & $0.80$ & $-$ & $0.002$ & $0.017$ & $0.014$ & $0.013$ \\ %updated 

    \cline{2-11}
    & Binary AB 
    & $7031.8/6993$
    & $5786.4018$ & $3.85$ & $145.56$ & $<1.58$ & $3.948$ & $25.065$ & $0.605$ & $19.426$ \\
    & & & $0.0005$ & $0.30$ & $11.15$ & $-$ & $0.003$ & $0.477$ & $0.088$ & $0.013$ \\ %updated 

    \cline{2-11}
    & Binary BC 
    & $7053.4/6993$
    & $5786.3960$ & $1.96$ & $290.79$ & $<0.92$ & $2.391$ & $26.068$ & $1.299$ & $19.402$ \\
    & & & $0.0004$ & $0.13$ & $18.34$ & $-$ & $0.002$ & $0.304$ & $0.059$ & $0.012$ \\ %updated 

    \cline{2-11}
    & Binary CD 
    & $7045.4/6993$
    & $5786.3987$ & $1.78$ & $320.38$ & $<0.85$ & $0.832$ & $26.122$ & $1.383$ & $19.406$ \\
    & & & $0.0005$ & $0.11$ & $19.94$ & $-$ & $0.002$ & $0.306$ & $0.057$ & $0.011$ \\ %updated 

    \cline{2-11}
    & {Binary DA}
    & $7018.7/6993$
    & $5786.3932$ & $5.28$ & $105.29$ & $<1.74$ & $5.566$ & $22.214$ & $0.198$ & $19.437$ \\
    & & & $0.0005$ & $0.34$ & $6.77$ & $-$ & $0.004$ & $0.914$ & $0.099$ & $0.013$ \\ %updated 
    
    \hline
    \hline
    \end{tabular}
    
    \begin{tablenotes}
        \centering
        \item{NOTE. } 
        ${\rm HJD}^{\prime} = {\rm HJD} - 2450000$. The parameters are reported with their $1\sigma$ uncertainties, while the upper limits on $\rho$ corresponds to $3\sigma$. The values of $t_0$ and $u_0$ are referenced to the magnification center, using the definition in \citetalias{Three_planet_candidates}. The source magnitude is calibrated to the OGLE-III photometric system. The best-fit solution is highlighted in bold. 
    \end{tablenotes}
    \label{OB110950_parm_2L1Sstatic}
\end{table*}

\section{Source Properties}\label{sec:source}
\begin{figure}
    \centering
    % \captionsetup{type=figure}
    \includegraphics[width=\columnwidth]{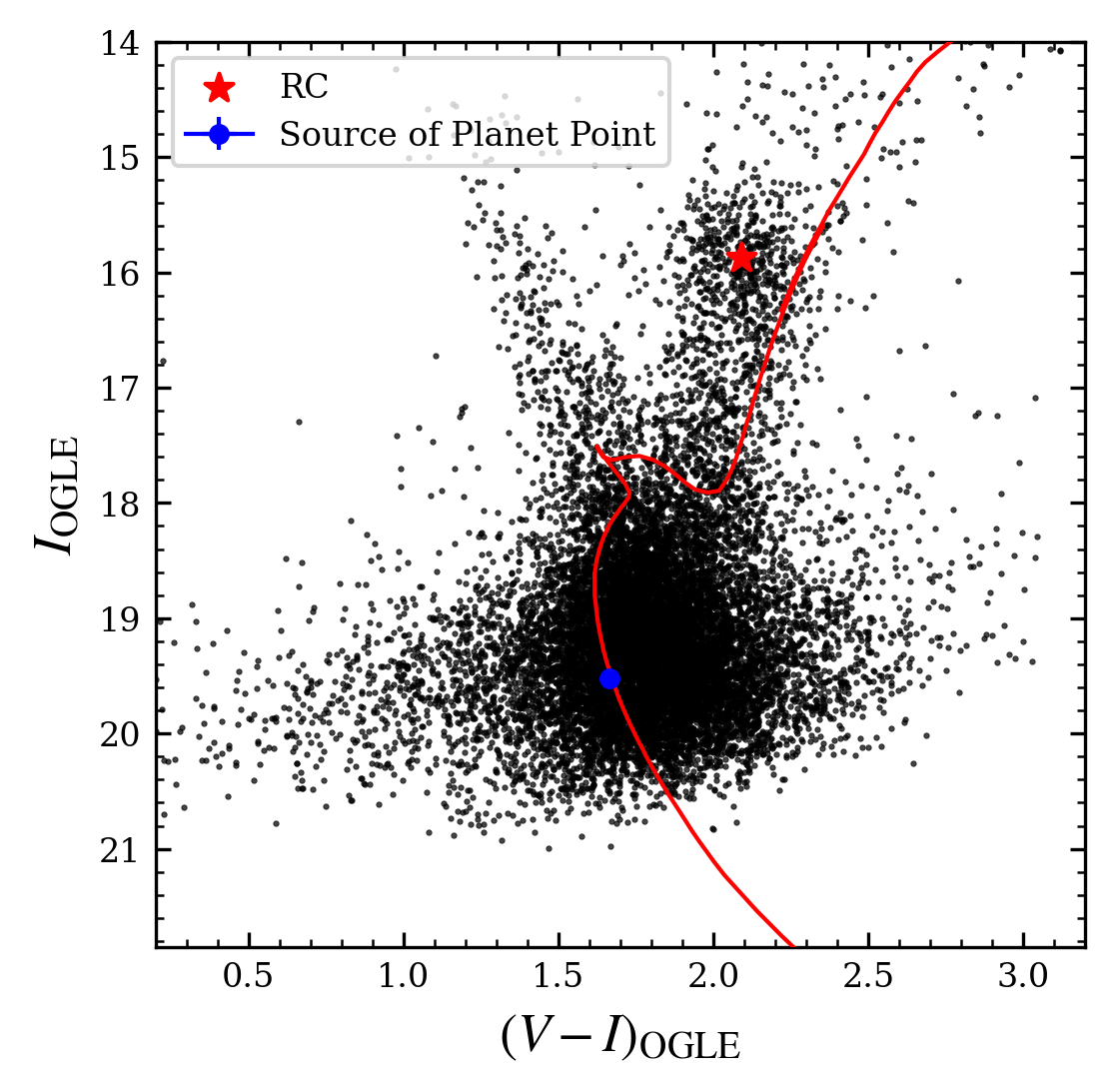}
    \caption{$V-I$ versus $I$ color-magnitude diagram of \event\ constructed from OGLE-III field stars \citep{OGLEIII} within a $2.5^\prime$ radius of the event. The red asterisk marks the centroid of the red clump (RC). The blue dot indicates the source position for the best-fit ``Planet Point'' solution, with its color derived from the OGLE-IV $V$- and $I$-band light curves. The red curve shows a PARSEC isochrone with an age of 2.4\,Gyr, a $\rm [M/H]$ of $+0.33$, and a distance of 9.2\,kpc, adopting the same extinction as the RC. The age and metallicity are taken from the spectroscopic measurements of \citet{Bensby2013}. The source properties inferred from the OGLE-IV light curves are consistent with these spectroscopic constraints. }
    \label{OB110950_cmd}
\end{figure}

\begin{figure}
    \centering
    % \captionsetup{type=figure}
    \includegraphics[width=\columnwidth]{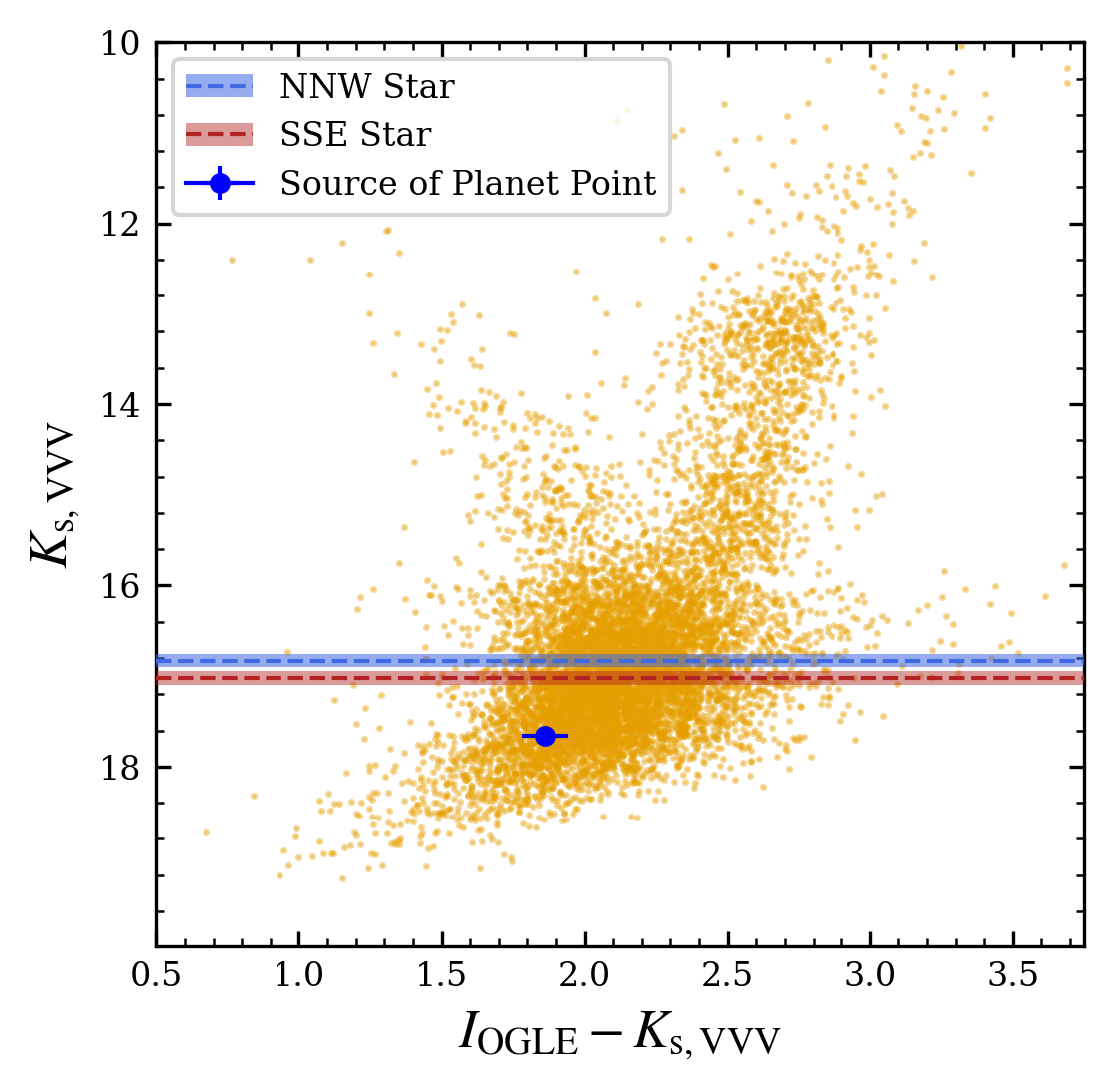}
    \caption{$I - K_{\rm s}$ versus $K_{\rm s}$ CMD constructed from OGLE-III and VVV stars within a $2.5^{\prime}$ radius of the event. The blue and red dashed lines indicate the magnitudes of the ``NNW'' and ``SSE'' stars identified in the Keck AO images, respectively, while the shaded bands indicate their $1\sigma$ uncertainties. The blue dot marks the position of the microlensed source star for the best-fit ``Planet Point'' solution. 
    }
    \label{VVV_cmd}
\end{figure}

We analyze the source properties in this section for three main reasons. First, the de-reddened source color and magnitude allow us to determine the angular source radius, $\theta_*$, which (for the case that $\rho$ is measured) can then be used to derive $\thetae$ and the geocentric lens-source relative proper motion via
\begin{equation}\label{equ:murel}
    \thetae = \frac{\theta_*}{\rho}, \quad
    \murelGeo = \frac{\thetae}{\te}.
\end{equation}
Different models may predict different values of $\murelGeo$, which can be compared with the heliocentric lens-source relative proper motion, $\mu_{\rm rel,H}$, which was measured from the Keck AO imaging presented by \citetalias{OB110950_AO}, after accounting for the transformation between the geocentric and heliocentric frames, to validate or rule out the models. 

Second, the de-reddened source color and magnitude enable us to infer the source apparent magnitude in the $K_{\rm s}$-band of the Keck AO imaging through a color-color relation, thereby inferring which of the two stars resolved by Keck AO corresponds to the source and which to the lens. Third, the de-reddened source color can be used to estimate the effective temperature, $T_{\rm eff}$, and thus to determine the limb-darkening coefficients adopted in the light-curve analysis.

\begin{table}
%[htb]
    \renewcommand\arraystretch{1.15}
    \centering
    % \caption{CMD parameters, 2L1S source properties and derived $\thetae$ and $\murel$ for \event}
    \caption{Derived $\theta_*$, $\thetae$, $\murelGeo$, and $K_{\rm s,Source}$ for \event}
    \begin{tabular}{c|c|r r r r}
    \hline
    \hline
    
    % \multicolumn{2}{c|}{Model} & 
    % $\theta_*$ ($\mu$as) &
    % $\thetae$ (mas) &
    % $\murel$ (${\rm mas\,yr^{-1}}$) \\
    
    % \multirow{2}{*}{\multicolumn{2}{c|}{Model}}  &
    % % $\theta_*$ ($\mu$as) &
    % % $\thetae$ (mas) &
    % \shortstack{$\theta_*$ \\ ($\mu$as)} &
    % \shortstack{$\thetae$ \\ (mas)} &
    % \shortstack{$\murel$ \\ (${\rm mas\,yr^{-1}}$)} \\
    
    % \hline

    \multicolumn{2}{c|}{Model} & $\theta_*$ & $\theta_{\rm E}$ & $\mu_{\rm rel, G}$ & $K_{\rm s,Source}$ \\
    % \cline{3-5}
    \multicolumn{2}{c|}{Unit} & $\mu$as & mas & mas\,yr$^{-1}$ & mag \\
    \hline

    \multirow{8}{*}{Close} 
    
    & {Planet Finite}
      & $0.70$ & $0.154$ & $0.83$ & $17.63$ \\
    & & $0.03$ & $0.007$ & $0.04$ & $0.07$ \\ 

    \cline{2-6}
    & {Planet Point}
      & $0.69$ & $>0.362$ & $>1.91$ & $17.66$ \\
    & & $0.03$ & $-$ & $-$ & $0.07$\\ 

    \cline{2-6}
    & Binary AB 
      & $0.72$ & $>0.269$ & $>1.53$ & $17.56$  \\
    & & $0.03$ & $-$ & $-$ & $0.07$  \\ 

    \cline{2-6}
    & {Binary BC}
      & $0.72$ & $>0.274$ & $>1.53$ & $17.58$  \\
    & & $0.03$ & $-$ & $-$ & $0.07$ \\

    \hline
    \multirow{12}{*}{Wide} 
    & {Planet Finite}
      & $0.70$ & $0.154$ & $0.83$ & $17.63$ \\
    & & $0.03$ & $0.007$ & $0.04$ & $0.07$ \\ 
         
    \cline{2-6}
    & {Planet Point}
      & $0.69$ & $>0.355$ & $>1.87$ & $17.66$  \\
    & & $0.03$ & $-$ & $-$ & $0.07$ \\      

    \cline{2-6}
    & Binary AB 
      & $0.72$ & $>0.458$ & $>1.15$ & $17.57$  \\
    & & $0.03$ & $-$ & $-$ & $0.07$ \\ 

    \cline{2-6}
    & Binary BC 
      & $0.73$ & $>0.795$ & $>1.00$ & $17.54$ \\
    & & $0.03$ & $-$ & $-$ & $0.07$ \\ 

    \cline{2-6}
    & Binary CD 
      & $0.73$ & $>0.859$ & $>0.98$ & $17.55$  \\
    & & $0.03$ & $-$ & $-$ & $0.07$ \\ 
    
    \cline{2-6}
    & {Binary DA}
      & $0.72$ & $>0.414$ & $>1.43$ & $17.58$ \\
    & & $0.03$ & $-$ & $-$ & $0.07$ \\ 
  
    \hline
    \hline
    \end{tabular}
    
    \begin{tablenotes}
        \centering
        \item{NOTE. } 
        % ${\rm HJD}^{\prime} = {\rm HJD} - 2450000$. 
        The parameters are presented with their $1 \sigma$ uncertainties. For the non-detection parameters, the $3 \sigma$ lower limits are provided. 
        % $t_0$ and $u_0$ take the magnification center as the reference point. 
        % The source magnitude has been calibrated to the standard $I$-band magnitude using the OGLE-III star catalog. 
        % % The pairs of planetary and binary models favored by $\chi^2$ are highlighted in bold. 
        % The best-fit model is highlighted in bold. 
        % %origin, which is the mass center when $s<1$ and ? when $s>1$. 
        % The best planet and binary models in both ``Close'' and ``Wide'' regime are highlighted in boldface. (explain the name and positive log q in advance)
    \end{tablenotes}
    %\tablecomments{The upper limits on $\rho$ are at $3\sigma$.}The upper limit on $\rho$ is $3\sigma$. 
    \label{table_CMD_ob110950}
\end{table}

\subsection{Lens-Source Relative Proper Motion}
\label{subsec:relative_proper_motion}
Based on the source color information from the SMARTS-CTIO $V$ and $I$ data and the OGLE-III color-magnitude diagram (CMD), \citetalias{OB110950_AO} found a de-reddened source color of $(V-I)_{\rm S,0} = 0.88 \pm 0.09$, and the subsequent source properties (e.g., $\theta_*$) as well as $\murelGeo $ for different solutions are based on this value. For this event, a high-resolution spectrum was obtained using the Ultraviolet and Visual Echelle Spectrograph (UVES) on the Very Large Telescope (VLT) at a magnification of $A \sim 85$, i.e., close to the time of peak magnification, which showed that the source is a mildly evolved, metal-rich dwarf with an effective temperature of $T_{\rm eff} = 6130 \pm 120~{\rm K}$, a surface gravity of $\log g = 4.20 \pm 0.15$, a stellar mass of $M = 1.27^{+0.13}_{-0.07}\,M_\odot$, a metallicity of $[{\rm Fe/H}] = +0.33 \pm 0.10$, and an age of $2.4^{+1.2}_{-0.8}~{\rm Gyr}$ \citep{Bensby2013,Bensby2017} \footnote{The full set of parameters of the source star is available in Table~A.2 of \citet{Bensby2017} via VizieR: \url{https://vizier.cds.unistra.fr/viz-bin/VizieR-5?-ref=VIZ693fb9963b6bf&-out.add=.&-source=J/A+A/605/A89/tablea2&recno=7}}. This yields the intrinsic source color of $(V-I)_{\rm S,0} = 0.60^{+0.04}_{-0.03}$, which is in strong tension with the CMD based color measurement of \citetalias{OB110950_AO}.

To resolve this discrepancy, we include the OGLE-IV $V$-band data in the analysis and estimate the apparent source color as $(V-I)_{\rm S} = 1.663 \pm 0.003$ based on a regression of OGLE-IV $V$ against $I$ flux as a function of magnification, followed by a calibration to the OGLE-III photometric system. We further analyze the OGLE-III $V-I$ versus $I$ CMD using stars within $2.5^\prime$ of the source position. The resulting CMD is shown in Figure~\ref{OB110950_cmd}. Applying the method of \citet{Nataf2013}, we measure the observed centroid of the red clump (RC) to be $(V-I, I)_{\rm RC} = (2.090 \pm 0.007, 15.872 \pm 0.031)$. 
% We adopt the de-reddened color and magnitude of the RC as $(V-I, I)_{\rm RC,0} = (1.06 \pm 0.03, 14.55 \pm 0.04)$ from \citet{Bensby2013} and \citet{Nataf2013}, respectively, evaluated at the Galactic coordinates of the event. 
We adopt the de-reddened color and magnitude of the RC as $(V-I,I)_{\rm RC,0} = (1.06 \pm 0.03,\, 14.55 \pm 0.04)$, taking $(V-I)_{\rm RC,0}$ from \citet{Bensby2013} and $I_{\rm RC,0}$ from Table~1 of \citet{Nataf2013}, evaluated at the event longitude $l=-1.93^\circ$. 
Following
\begin{equation}\label{equ:extinction1}
[E(V-I), A_I] = (V-I, I)_{\rm RC} - (V-I, I)_{\rm RC,0},
\end{equation}
we derive the reddening and extinction toward the event to be $E(V-I) = 1.03 \pm 0.03$ and $A_I = 1.32 \pm 0.05$. 

The source apparent magnitude $I_{\rm S}$, which differs among the solutions, is listed in Table~\ref{OB110950_parm_2L1Sstatic}. Applying
\begin{equation}\label{equ:extinction2}
(V-I, I)_{\rm S,0} = (V-I, I)_{\rm S} - [E(V-I), A_I],
\end{equation}
we obtain a de-reddened source color of $(V-I)_{\rm S,0} = 0.63 \pm 0.03$, which is consistent with the value inferred from the VLT spectrum. A comparison between our CMD analysis and that of \citetalias{OB110950_AO} indicates that the primary difference arises from the adopted reddening value: we obtain $E(V-I) = 1.03 \pm 0.03$, whereas \citetalias{OB110950_AO} adopted $E(V-I) = 0.83$. The reddening value used by \citetalias{OB110950_AO} implies an extinction law of $A_I/E(V-I) = 1.6$, which is higher than typically observed values of $A_I/E(V-I) = 1.0$--$1.4$ toward the Galactic bulge (Figure~6 of \citealt{Nataf2013}). Because \citetalias{OB110950_AO} adopted a redder intrinsic source color, the corresponding values of $\theta_*$, $\thetae$ and $\murelGeo$ are overestimated, and the inferred source flux in the $K_{\rm s}$-band is also overestimated. 

Using the color-surface brightness relation of \citet{Adams2018},
\begin{equation}\label{equ:theta*}
\log (2\theta_*) = 0.378\,(V-I)_{\rm S,0} + 0.542 - 0.2\,I_{\rm S,0},
\end{equation}
we compute the angular source radius $\theta_*$. $\thetae$ and $\murelGeo$ are then derived using Equation~(\ref{equ:murel}). 
The inferred $\theta_*$, $\thetae$, and $\murelGeo$ for different solutions are listed in Table~\ref{table_CMD_ob110950}. 

Because the inferred lens-source relative proper motion for the ``Planet Finite'' solutions, $\mu_{\rm rel,G} = 0.83 \pm 0.04~{\rm mas\,yr^{-1}}$, is even lower than the value reported by \citetalias{OB110950_AO}, it is in stronger tension with the $\mu_{\rm rel,G} \simeq 4~{\rm mas\,yr^{-1}}$ obtained by converting the Keck AO heliocentric measurement of \citetalias{OB110950_AO} to the geocentric frame. We therefore exclude the ``Planet Finite'' solutions. In contrast, the ``Planet Point'' and ``Binary'' solutions yield only lower limits on the relative proper motion, $\mu_{\rm rel,G} \gtrsim 1$--$2~{\rm mas\,yr^{-1}}$. They are therefore not excluded by the relative proper motion analysis. In addition, among the ``Binary'' solutions, the four ``Wide Binary'' are excluded because they would show a resolvable companion that is not seen in the Keck AO imaging (See Section~6 of \citetalias{OB110950_AO}). Thus, in the next section, we only derive the lens physical parameters for the ``Planet Point'' and ``Close Binary'' solutions.

\subsection{Source and Lens Star Identification}
\label{subsec:source_lens_identification}
To determine which of the two stars resolved in the Keck AO images by \citetalias{OB110950_AO} corresponds to the source and which to the lens, we first estimate the source $K_{\rm s}$-band magnitude, $K_{\rm s,Source}$, using a color-color relation.
From the CMD analysis in Section~\ref{subsec:relative_proper_motion}, we derive a de-reddened source color of $(V-I)_{\rm S,0} = 0.63 \pm 0.03$. Using the color-color relation of \cite{color_color_relation_Pecaut2013}, this corresponds to $(I-K_{\rm s})_{\rm S,0} = 0.74 \pm 0.04$. Employing the  relation of \citet{Kenyon1995} yields a consistent value within $1\sigma$. To estimate the extinction in the $K_{\rm s}$ band, we construct an $I_{\rm OGLE} - K_{\rm s,VVV}$ versus $K_{\rm s,VVV}$ CMD using nearby field stars from the OGLE-III survey and the VISTA Variables in the Via Lactea (VVV) survey \citep{VVV, VVV_catlog_2019}; see Figure~\ref{VVV_cmd}. Applying the same method as described in Section~\ref{subsec:relative_proper_motion}, we derive a $K_{\rm s}$-band extinction of $A_{K_{\rm s}} = 0.20 \pm 0.04$, consistent with the extinction law relating $A_{K_{\rm s}}$ and $A_I$ given by \citet{Nataf2016}.
For the ``Planet Point'' solutions, the inferred de-reddened source magnitude is $I_{\rm S,0} = 18.20 \pm 0.05$, which implies
\begin{equation}
    K_{\rm s,Source} = I_{\rm S,0} - (I-K_{\rm s})_{\rm S,0} + A_{K_{\rm s}} = 17.66 \pm 0.07. 
\end{equation} 
Different solutions yield slightly different $I_{\rm S}$ and therefore slightly different $K_{\rm s,Source}$, but the spread is only $\sim 0.1$~mag. The resulting $K_{\rm s,Source}$ values for different solutions are listed in Table~\ref{table_CMD_ob110950}. 

The Keck AO observations presented by \citetalias{OB110950_AO} resolve two stars at the event position, with $K_{\rm SSE} = 17.02 \pm 0.08$ and $K_{\rm NNW} = 16.83 \pm 0.07$, where SSE and NNW denote the south-southeast and north-northwest stars, respectively. 
A clear tension arises because our predicted source $K_{\rm s}$-band magnitude is fainter than both the SSE and NNW stars by $\sim 0.6$~mag and $\sim 0.8$~mag for the ``Planet Point'' solutions and $\sim 0.5$~mag and $\sim 0.7$~mag for the ``Close Binary'' solutions. Therefore, in either configuration, the source star must have at least one unresolved companion contributing comparable $K_{\rm s}$-band flux. 
We also note that our derived $K_{\rm s,Source}$ is $\sim 0.5$~mag fainter than the value reported by \citetalias{OB110950_AO}. 
This difference is primarily due to their adoption of a redder intrinsic source color, which leads to an overestimate of the source $K_{\rm s}$-band flux, as discussed in Section~\ref{subsec:relative_proper_motion}. 

We next assess whether a binary source interpretation is compatible with both the microlensing light curve and the Keck AO images. As an illustrative example, we consider a second source with a relatively large impact parameter, $u_{0,2} \sim 5$. In this case, the second source would remain only weakly magnified and make a negligible contribution to the light curve, so that, together with the highly magnified primary source, the event could be well fitted by a single source model. 
Using the values of $\thetae$ derived in Section~\ref{sec:lens}, such a configuration corresponds to an angular separation of $\sim 4$~mas between the two sources. This is much smaller than the Keck AO point spread function (PSF) full width at half maximum (FWHM) of $\sim 67$~mas reported by \citetalias{OB110950_AO}, and hence the two sources would be unresolved in the AO images. 
We note that although this estimate is based on a single representative value of $u_{0,2}$, any configuration in which the second source passes several $\thetae$ away from the lens would similarly satisfy both the light curve and imaging constraints. At the distance of the source, an angular separation of $\sim 4$~mas corresponds to a projected separation of $\sim 40$~au. For a $\sim 1.3~M_\odot$ source (as inferred from the VLT spectrum) with a companion of comparable mass, this separation implies an orbital period $\gtrsim 150$~yr, so any xallarap effect \citep{HardyWalker1995} on the microlensing light curve can safely be neglected. 

% To examine whether a binary source interpretation is allowed by the microlensing light curve and the Keck AO observation, we (do some calculation here/estimate) and show that there is large freedom to not violate both constraints. 
% Assuming a relatively large second source's impact parameter, e.g. u0,2 ~ 4, which will have no additional signature on the light curve, and considering the highly amplified primary source, the resulting light curve can be well fitted with a single source model. 
% And the u0,2 ~ 4 translates to a ~3 mas angular separation between the binary sources in the AO image using the thetaE values derived in Section 4, which is too small compared to a full width at half-maximum (FWHM) of ~67 mas for the Keck AO observation and thus will also be unresolvable from AO image. 
% A similar case is seen in \citet{OB161195_AO}, where the binary source interpretation is allowed by both the light curve and AO observation. 
% And such a binary source corresponds to a projected separation of ~30 au, and a >~120 year orbital period, thus the xallarap effect can be negelected. 

Because the source star could be located at the position of either Keck star, both the NNW and SSE stars are, in principle, viable lens candidates. Using the Galactic model of \citet{Koshimoto2021} and the fact that the direction of $\boldsymbol{\mu}_{\rm rel} \equiv \boldsymbol{\mu}_{\rm L}-\boldsymbol{\mu}_{\rm S}$ flips when swapping the lens and source, \citetalias{OB110950_AO} found that the NNW star is 3.79 times more likely to be the lens than the SSE star, although the latter possibility cannot be completely excluded. We therefore derive the lens physical parameters for two scenarios, one in which the NNW star is the lens and one in which the SSE star is the lens. However, the final inferred lens properties are only weakly sensitive to this choice, because the two stars have very similar $K_{\rm s}$-band fluxes and will therefore yield similar lens masses and distances.

\section{Lens Properties}\label{sec:lens}
\begin{figure*}
    %\centering
    \includegraphics[width=1.00\textwidth]{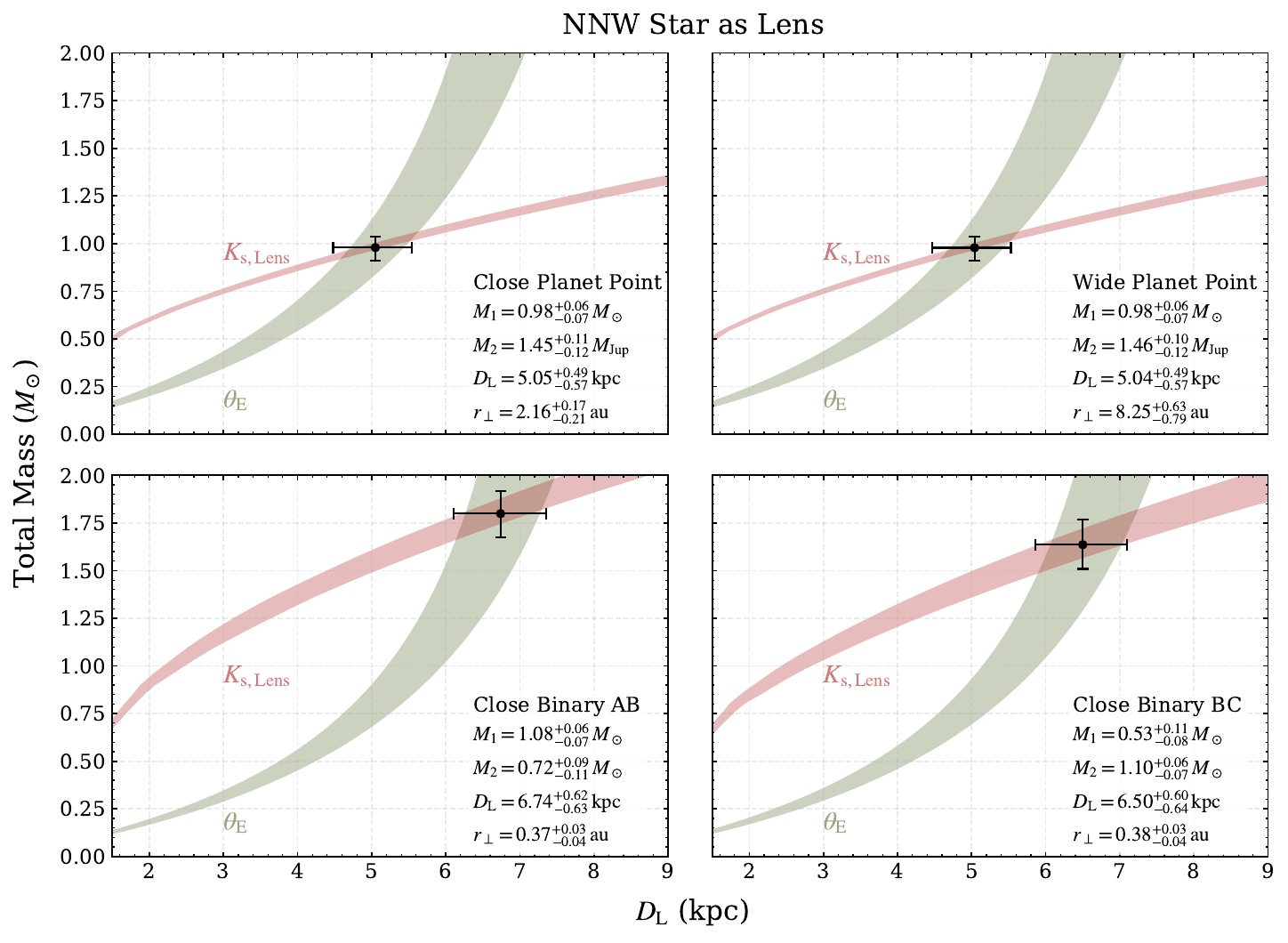}
    \caption{
    Constraints on the lens distance $D_{\rm L}$ and total lens mass $M_1+M_2$ for the four viable 2L1S solutions of \event, assuming that the NNW star is the lens. 
    In each panel, the green band shows the $\theta_{\rm E}$ constraint derived from $\te$ and the Keck AO measurement of the lens-source relative proper motion, accounting for the uncertainties in $\theta_{\rm E}$ and $D_{\rm S}$. 
    The red band shows the constraint from the Keck $K_{\rm s}$-band lens flux using the empirical mass--luminosity relation, accounting for the uncertainties in the lens-flux measurement, extinction, and binary mass ratio. 
    For the two ``Close Binary'' solutions, the Keck lens flux is interpreted as the combined flux of the two stellar components. 
    The black point with $1\sigma$ error bars indicates the median values of $(D_{\rm L}, M_1+M_2)$ and their uncertainties from the forward modeling. }
\label{fig:NNW}
\end{figure*}

\begin{figure*}
    %\centering
    \includegraphics[width=1.00\textwidth]{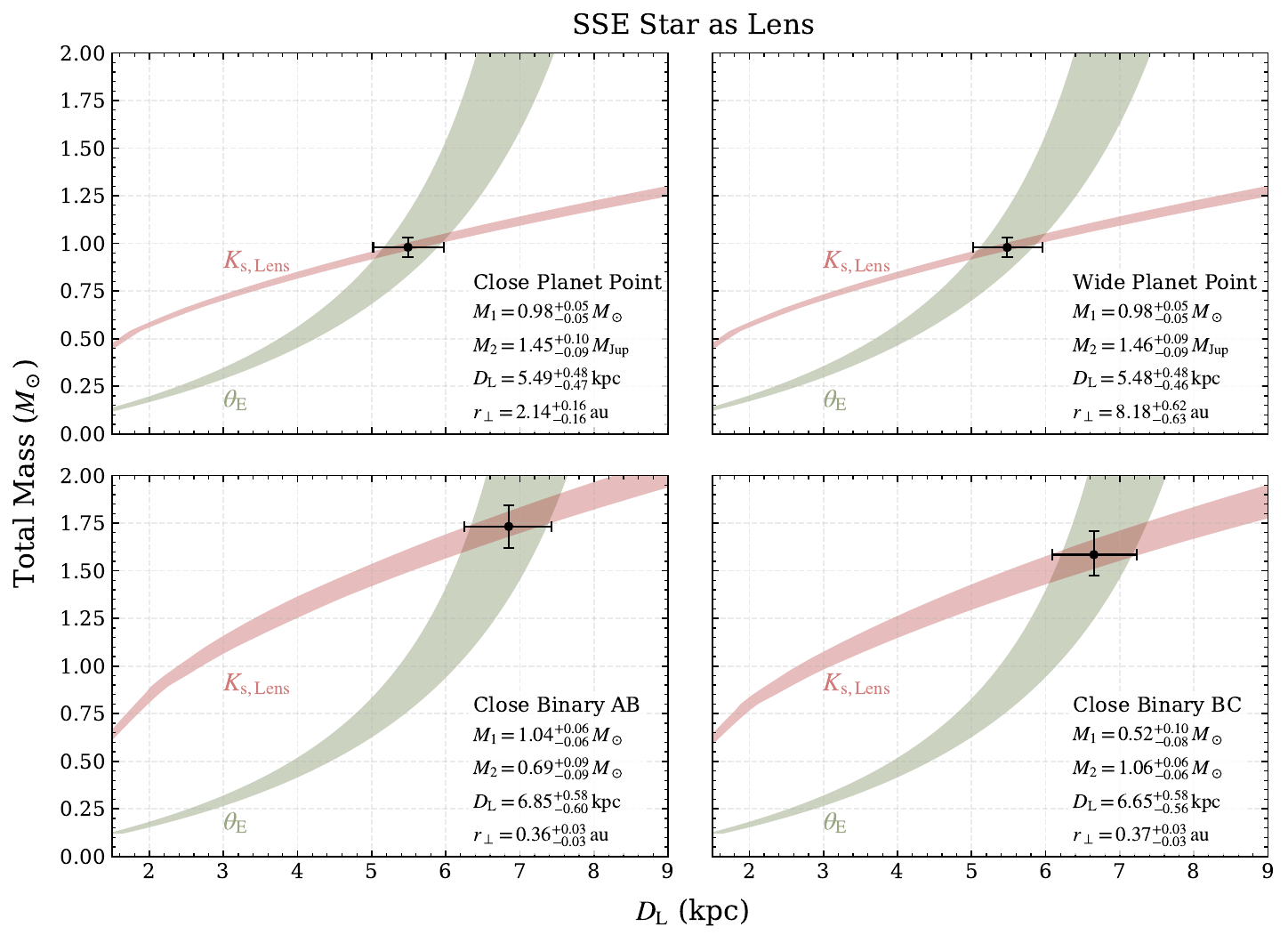}
    \caption{
    Same as Figure~\ref{fig:NNW}, but assuming that the SSE star is the lens. Because the SSE star is fainter than the NNW star and the SSE-star-as-lens scenario corresponds to a smaller $\mu_{\rm rel,G}$, hence a smaller $\theta_{\rm E}$, the allowed solutions shift to larger $D_{\rm L}$. 
    }
\label{fig:SSE}
\end{figure*}

\begin{table*}
%[htb]
    \renewcommand\arraystretch{1.8}
    \centering
    \caption{Lens Properties for~\event
    % ~(NNW Star as Lens, $K_{\rm s, NNW} = 16.83 \pm 0.07$)
    }

    \begin{tabular}{c|c|ccccccc}
    \hline
    \hline
    \multirow{2}{*}{Lens} & \multirow{2}{*}{Solution}& $D_{\rm S}$ & $D_{\rm L}$  
  & $M_{1}$ & $M_{2}$  
  & $r_{\perp}$  & $\mu_{\rm rel, G}$  
  & $\thetae$ \\ 
  % \cline{3-9} \cline{10-12}
  
   & & (kpc) & (kpc) 
  & ($M_{\odot}$) &   
  & (au) & (${\rm mas\,yr^{-1}}$) 
  & (mas)\\ 
  \hline
    
    % \hline
    \multirow{4}{*}{NNW Star} 
    & Close Planet Point
    & $9.20^{+0.88}_{-0.89}$
    & $5.05^{+0.49}_{-0.57}$
    & $0.98^{+0.06}_{-0.07}$
    & $1.45^{+0.11}_{-0.12}$ $M_{\rm Jup}$
    & $2.16^{+0.17}_{-0.21}$
    & $4.44^{+0.24}_{-0.19}$
    & $0.84^{+0.05}_{-0.04}$ \\ %
    
    \cline{2-9}
     & Wide Planet Point
    & $9.21^{+0.88}_{-0.90}$
    & $5.04^{+0.49}_{-0.57}$
    & $0.98^{+0.06}_{-0.07}$
    & $1.46^{+0.10}_{-0.12}$ $M_{\rm Jup}$
    & $8.25^{+0.63}_{-0.79}$
    & $4.44^{+0.24}_{-0.19}$
    & $0.84^{+0.05}_{-0.04}$ \\ % 

    \cline{2-9}
     & Close Binary AB
    & $9.23^{+0.89}_{-0.91}$
    & $6.74^{+0.62}_{-0.63}$
    & $1.08^{+0.06}_{-0.07}$
    & $0.72^{+0.09}_{-0.11}$ $M_{\odot}$
    & $0.37^{+0.03}_{-0.04}$
    & $4.31^{+0.18}_{-0.17}$ 
    & $0.76^{+0.03}_{-0.03}$ \\ % 
    
    \cline{2-9}
     & Close Binary BC
    & $9.23^{+0.86}_{-0.93}$
    & $6.50^{+0.60}_{-0.64}$
    & $0.53^{+0.11}_{-0.08}$
    & $1.10^{+0.06}_{-0.07}$ $M_{\odot}$
    & $0.38^{+0.03}_{-0.04}$
    & $4.33^{+0.19}_{-0.18}$ 
    & $0.77^{+0.04}_{-0.03}$ \\ % 7628.138016907755

    \hline
    \multirow{4}{*}{SSE Star} 
    & Close Planet Point
    & $9.21^{+0.90}_{-0.88}$
    & $5.49^{+0.48}_{-0.47}$
    & $0.98^{+0.05}_{-0.05}$
    & $1.45^{+0.10}_{-0.09}$ $M_{\rm Jup}$
    & $2.14^{+0.16}_{-0.16}$
    & $4.04^{+0.15}_{-0.15}$
    & $0.76^{+0.03}_{-0.03}$ \\ % 
    
    \cline{2-9}
     & Wide Planet Point
    & $9.22^{+0.90}_{-0.88}$
    & $5.48^{+0.48}_{-0.46}$
    & $0.98^{+0.05}_{-0.05}$
    & $1.46^{+0.09}_{-0.09}$ $M_{\rm Jup}$
    & $8.18^{+0.62}_{-0.63}$
    & $4.04^{+0.15}_{-0.15}$
    & $0.77^{+0.03}_{-0.03}$ \\ % 

    \cline{2-9}
     & Close Binary AB
    & $9.24^{+0.89}_{-0.91}$
    & $6.85^{+0.58}_{-0.60}$
    & $1.04^{+0.06}_{-0.06}$
    & $0.69^{+0.09}_{-0.09}$ $M_{\odot}$
    & $0.36^{+0.03}_{-0.03}$
    & $4.12^{+0.15}_{-0.15}$
    & $0.73^{+0.03}_{-0.03}$ \\ % 
    
    \cline{2-9}
     & Close Binary BC
    & $9.24^{+0.86}_{-0.88}$
    & $6.65^{+0.58}_{-0.56}$
    & $0.52^{+0.10}_{-0.08}$
    & $1.06^{+0.06}_{-0.06}$ $M_{\odot}$
    & $0.37^{+0.03}_{-0.03}$
    & $4.11^{+0.15}_{-0.15}$
    & $0.73^{+0.03}_{-0.03}$ \\ % 

    \hline
    \hline
    \end{tabular}
    
    %\begin{tablenotes}
     %   \centering
      %  \item{NOTE. } 
       % The parameters are presented with their $1 \sigma$ uncertainties. 
        % \textcolor{red}{Jiyuan: new code just finished. now more consistent with T22 than before. with informatvie prior, Dl=7.0 kpc; with uniform prior, Dl=6.5 kpc; so the difference is significant.  }
    %\end{tablenotes}
    \label{table_lens_property}
\end{table*}

To determine the lens mass $M_{\rm L}$ and distance $D_{\rm L}$, two independent mass-distance relations are required. For this event, one constraint is provided by the Keck AO observations, which yield a measurement of the lens brightness. A second relation arises from the connection between $M_{\rm L}$, $D_{\rm L}$, the angular Einstein radius $\thetae$, and the source distance $D_{\rm S}$, given by
% \begin{equation}\label{eq:MLDL}
% % M_{\rm L} = \frac{\theta_{\rm E}^2 D_{\rm L} D_{\rm S}}{\kappa (D_{\rm S} - D_{\rm L})},
% M_{\rm L} = \frac{\theta_{\rm E}^2 D_{\rm L} D_{\rm S}}{\kappa (D_{\rm S} - D_{\rm L})},
% \end{equation}
% where $\kappa \equiv 4G/(c^2\,\mathrm{au}) \simeq 8.144~\mathrm{mas}\,M_\odot^{-1}$.
\begin{equation}\label{eq:MLDL}
M_{\rm L} \;=\; \frac{\theta_{\rm E}^2}{\kappa\,\pi_{\rm rel}},
\qquad
\pi_{\rm rel} \;=\; \mathrm{au}\left(\frac{1}{D_{\rm L}}-\frac{1}{D_{\rm S}}\right),
\end{equation}
where $\kappa \equiv 4G/(c^2\,\mathrm{au}) \simeq 8.144~\mathrm{mas}\,M_\odot^{-1}$, and $\pi_{\rm rel}$ is the lens-source relative parallax.

For this event, $\thetae$ is obtained by combining the lens-source relative proper motion with the event time scale via Equation~(\ref{equ:murel}). However, the uncertainty in the source distance inferred from spectroscopy is relatively large, $D_{\rm S} = 9.7^{+2.9}_{-1.4}\,\mathrm{kpc}$. Moreover, accurately propagating the uncertainties in all observables is most naturally done within a forward-modeling framework, which we adopt in the following analysis.

For each solution, we simulate $5\times10^{7}$ mock microlensing events. The source distances, $D_{\rm S}$, are drawn from the Galactic model of \citet{Yang2021_GalacticModel} and reweighted by the cumulative stellar density in front of the source to account for the higher microlensing probability of sources with more foreground stars, while lens masses, $M_{\rm L}$, and distances, $D_{\rm L}$, are sampled from uniform distributions. These quantities are then used to compute $\thetae$ via Equation~(\ref{eq:MLDL}). We adopt an extinction profile along the line of sight based on the exponential dust distribution used by \citetalias{OB110950_AO}. Together with the empirical mass--luminosity relations from \citet{Henry1993, Henry1999, Delfosse2000} used by \citetalias{OB110950_AO}, this allows us to derive the lens $K_{\rm s}$-band magnitude, $K_{\rm s,Lens}$. The impact of different modeling assumptions on the inferred lens properties is discussed in Section~\ref{sec:dis}.

The heliocentric lens-source relative proper motion, $\mu_{\rm rel,H}$, is constrained by the Keck AO measurement. For each simulated event, we draw $\boldsymbol{\mu}_{\rm rel,H}$ from the two-dimensional Gaussian likelihood defined by this measurement and then convert it to the geocentric frame via
% The heliocentric lens-source relative proper motion, $\mu_{\rm rel,H}$, is taken from the Keck AO measurement. For each simulated event, the corresponding geocentric relative proper motion, $\mu_{\rm rel,G}$, is calculated via
\begin{equation}
\boldsymbol{\mu}_{\rm rel,G} = \boldsymbol{\mu}_{\rm rel,H} - \frac{\boldsymbol{v}_{\oplus, \perp}\pi_{\rm rel}}{\rm au},
\end{equation}
where $\boldsymbol{v}_{\oplus, \perp}$ is Earth's projected velocity relative to the Sun at the time of peak magnification. The event time scale is then obtained using Equation~(\ref{equ:murel}). Finally, each simulated event is weighted by
\begin{equation}\label{equ: weight}
w = p(\te)\times p(K_{\rm s,Lens}),
\end{equation}
where $p(\te)$ is the likelihood function for $\te$ derived from the light-curve analysis, and $p(K_{\rm s,Lens})$ is the likelihood function for $K_{\rm s,Lens}$ derived from the Keck AO observations. 

Figures~\ref{fig:NNW} and \ref{fig:SSE} show the constraints on $M_{\rm L}$ and $D_{\rm L}$ derived from the lens $K_{\rm s}$-band magnitude and $\thetae$, assuming the NNW star to be the lens and the SSE star to be the lens, respectively. Table~\ref{table_lens_property} summarizes the inferred physical parameters of the lens system, including $D_{\rm S}$, $D_{\rm L}$, $\mu_{\rm rel,G}$, and the component masses of the lens system, $M_1$ and $M_2$, given by
\begin{equation}
M_1 = \frac{M_{\rm L}}{1+q}, \qquad
M_2 = \frac{qM_{\rm L}}{1+q},
\end{equation}
as well as the projected separation between the two lens components, $r_\perp$, and $\thetae$.

For the ``Planet Point'' solutions, the host is a solar-mass star and the companion is a super-Jupiter. Adopting a snow-line scaling of $r_{\rm SL} = 2.7 \left(\frac{M_1}{M_\odot}\right)\,{\rm au}$
\citep{snowline}, the super-Jupiter lies well beyond the snow line in the ``Wide'' topology, whereas in the ``Close'' topology it is probably located near the snow line. For the ``Close Binary'' solutions, the more massive component is an F/G-type star, while the less massive component is likely a K-type star for the ``AB'' solution and an M-type star for the ``BC'' solution. The projected separation is $\sim 0.4$ au. Compared to the ``Planet Point'' solutions, the ``Close Binary'' solutions favor a slightly more massive primary and a more distant lens system. This arises because the ``Close Binary'' solutions have a shorter event time scale and therefore a smaller $\thetae$. In addition, the $\thetae$ of the more massive component is further reduced by a factor of $1/\sqrt{1+q}$. Because the lens flux is dominated by the more massive component, a smaller $\thetae$ implies a higher lens mass and a larger distance for a given observed luminosity. 
In addition, when the NNW star (which is $\sim 0.2$~mag brighter) is assumed to be the lens, the larger derived $\mu_{\rm rel,G}$ implies a larger $\theta_{\rm E}$, which in turn drives the solution toward a lens system that is closer to the Sun.

\section{Discussion}\label{sec:dis}

\subsection{Implications for Roman Microlensing}

The upcoming Nancy Grace Roman Space Telescope is expected to detect thousands of bound and free-floating planets through microlensing \citep{Spergel2015,MatthewWFIRSTI,Johnson2020}. One of Roman's key advantages over ground-based microlensing surveys is its ability to resolve the lens and source for many events via multi-year, high-resolution imaging. Our discovery of new ``Planet Point'' solutions for \event, arising from the newly identified ``Planet Finite/Planet Point'' degeneracy and retaining consistency with the relative proper motion constraint from the Keck AO observations, indicates that the ``Planet/Binary'' degeneracy may still introduce substantial ambiguities in some Roman planetary events. Moreover, the fact that the ``Planet Point'' solutions for \event\ were not recognized in previous analyses highlights the need for caution when interpreting events affected by the ``Planet/Binary'' degeneracy. 

Our revised analysis of the source properties reveals significant additional light at the source position, indicating that the source star has at least one companion. This represents the second microlensing event for which Keck AO imaging has shown that the source has a companion, with the other case being OGLE-2016-BLG-1195 \citep{OB161195,OB161195_MOA,OB161195_Gould,OB161195_AO}. Current estimates of Roman's capability to resolve the source and lens and to characterize the lens do not account for contamination from companions to the source or lens, nor from ambient field stars \citep{RomanIV}. Future analyses of Roman data will therefore need to carefully consider such contamination. In particular, if companions to the lens star contribute a significant fraction of the flux at the lens position but were not detected in the microlensing light curve, assuming a single lens star would bias the inferred host properties toward a more massive and more distant lens thereby affecting the derived planetary mass function, the inferred dependence of planet properties on their host stars, and the Galactic distributions of microlensing planets.

\begin{figure}
    \centering
    % \captionsetup{type=figure}
    \includegraphics[width=\columnwidth]{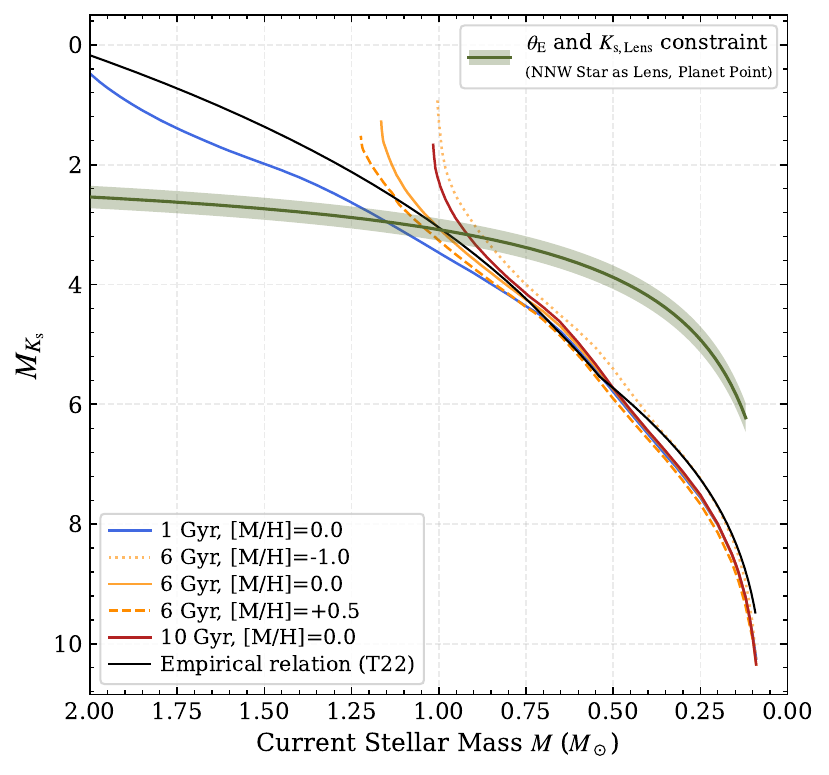}
    \caption{
    Comparison between the empirical and PARSEC mass--luminosity relations. 
    The black curve shows the empirical mass--luminosity relation adopted by \citetalias{OB110950_AO}. 
    Colored curves show PARSEC $K_{\rm s}$-band isochrones \citep{Bressan2012,Chen2014} for ages of 1, 6, and 10~Gyr and metallicities $\mathrm{[M/H]} = -1.0, 0.0,$ and $+0.5$. 
    The green band shows the mass--luminosity constraint inferred from $\thetae$ and Keck $K_{\rm s}$-band lens flux for the ``Planet Point'' solutions with the NNW star taken as the lens, accounting for the uncertainties in $\theta_{\rm E}$, $D_{\rm S}$, $K_{\rm s, Lens}$, and extinction. 
}
    \label{mass_luminosity_relation_plot}
\end{figure}

\subsection{Dependence on Mass--Luminosity Relation}

Previous treatments of the mass--luminosity relation in lens-flux analyses using Keck and the Hubble Space Telescope commonly adopt a single mass--luminosity relation (e.g., \citealt{OB120950}), implicitly assuming that the adopted relation is only weakly sensitive to stellar age and metallicity. 
This approximation is generally adequate for low-mass main-sequence lenses, for which the near-infrared mass--luminosity relation is nearly age/metallicity independent. 
However, for lenses with $M_{\rm L}\gtrsim 1\,M_\odot$, this assumption can break down as the lens approaches the turnoff/subgiant regime.

In this work, following \citetalias{OB110950_AO}, we adopt a single empirical mass--luminosity relation \citep{Henry1993,Henry1999,Delfosse2000}, which does not account for the uncertainties in the mass--luminosity relation arising from variations in stellar age and metallicity. Figure~\ref{mass_luminosity_relation_plot} compares this empirical relation to five representative isochrones from the PAdova and tRieste Stellar Evolution Code (PARSEC) project \citep{Bressan2012,Chen2014}, together with the mass--luminosity constraint inferred from $\thetae$ and Keck $K_{\rm s}$-band lens flux. 
The empirical relation intersects the PARSEC isochrone with an age of $6~{\rm Gyr}$ and a metallicity of $[{\rm M/H}] = 0$ near the mass favored by the ``Planet Point'' solutions. Varying the age from $1~{\rm Gyr}$ to $10~{\rm Gyr}$ changes the inferred lens mass by $\sim 0.23~M_{\odot}$, which is larger than the reported $1\sigma$ uncertainty of $\sim 0.07~M_{\odot}$ listed in Table~\ref{table_lens_property}. Similarly, varying the metallicity from $[{\rm M/H}] = -1.0$ to $+0.5$ results in a change in the inferred lens mass of $\sim 0.18~M_{\odot}$, again exceeding the uncertainty reported in Table~\ref{table_lens_property}. For the ``Close Binary'' solutions, the primary star alone is already more massive than the host in the ``Planet Point'' solutions. As a result, the same range of variations in stellar age and metallicity produce even larger shifts in the inferred lens mass. We therefore conclude that, for this event, the uncertainties in the lens properties are likely underestimated due to the simplified treatment of the mass--luminosity relations. 

On the other hand, in the low-mass regime ($\lesssim 0.8~M_{\odot}$), the uncertainties in the mass--luminosity relation induced by variations in stellar age and metallicity are smaller, but may still be comparable to those arising from the $\thetae$ constraint and the lens flux measurement.

Besides adopting a single mass--luminosity relation, some previous studies adopt the weighted isochrone distributions introduced by \citet{MB11291}; however, these works do not explicitly compare the adopted age and metallicity distributions with observations, nor do they clearly describe how the weights are derived. The recently published code \texttt{pyLIMASS} \citep{pylimass} incorporates multiple isochrones but assigns equal weight to each isochrone. In addition, current treatments of the foreground extinction distribution often rely on simple analytic prescriptions (e.g., \citetalias{OB110950_AO}).

With the Gaia satellite \citep{Gaia2016AA,Gaia2018AA}, multi-band photometric surveys, and extensive spectroscopic campaigns, the Galactic distributions of stellar ages and metallicities (e.g., \citealt{Anders2023}), as well as the three-dimensional extinction distribution (e.g., \citealt{Green2019}), are now much better constrained. With Roman scheduled for launch in 2026/2027, it is therefore timely to develop a lens-flux analysis framework that incorporates realistically weighted isochrone distributions and improved models of foreground extinction. We are currently developing such a framework and will present it in a forthcoming work (Zhang et al., in preparation).

\acknowledgments
J.Z., W.Z., H.Y. and S.M. acknowledge support by the National Natural Science Foundation of China (Grant No. 12133005). The OGLE project has received funding from the Polish National Science Centre grant OPUS 2024/55/B/ST9/00447 awarded to A.U.. H.Y. acknowledges support by the China Postdoctoral Science Foundation (No. 2024M762938). This work is part of the ET space mission which is funded by the China's Space Origins Exploration Program.

\bibliography{Zang.bib}

\end{CJK*}
\end{document}